\documentclass{PoS}%
\usepackage{amsmath}
\usepackage{amsfonts}
\usepackage{amssymb}
\usepackage{color}
\usepackage{bm}
\ifx\msidvipdf\msiundefined
\usepackage{graphicx}
\else%
\usepackage[dvipdfm]{graphicx}
\fi
\providecommand{\U}[1]{\protect\rule{.1in}{.1in}}
\providecommand{\U}[1]{\protect\rule{.1in}{.1in}}
\providecommand{\U}[1]{\protect\rule{.1in}{.1in}}

\title{%
\begin{flushright}
\begin{minipage}{0.3\textwidth} \normalsize
KEK Preprint 2018-81 \\  CHIBA-EP-229 \\
\end{minipage} \end{flushright} %
Confinement/deconfinement phase transition in SU(3) Yang-Mills theory and non-Abelian dual Meissner effect}

\ShortTitle{Confinement/deconfinement phase transition in SU(3) Yang-Mills theory and ... }

\author{\speaker{Akihiro Shibata}\\
    Computing Research Center, High Energy Accelerator Research Organization (KEK) \\
    SOKENDAI (The Graduate University for Advanced Studies),  Tuskuba 305-0801, Japan\\
    E-mail: \email{Akihiro.shibata@kek.jp}}

\author{Seikou Kato\\
        Oyama National College of Technology, Oyama 323-0806, Japan\\
        E-mail: \email{skato@oyama-ct.ac.jp}}

\author{Kei-Ichi Kondo\\
        Department of Physics, Graduate School of Science and Engineering, Chiba University \\
        Department of Physics, Graduate School of Science, Chiba University, Chiba 263-8522, Japan\\
        E-mail: \email{kondok@faculty.chiba-u.jp}}

\abstract{%
The dual superconductivity is a promising mechanism of quark confinement. In the preceding works, 
we have given a non-Abelian dual superconductivity picture for quark confinement, and demonstrated the numerical evidences on the lattice.
\\
In this talk, we focus on the the confinement and deconfinement phase transition at finite temperature in view of the dual superconductivity.
 By using our new formulation of lattice Yang-Mills theory and numerical simulations on the lattice, 
we extract the dominant mode for confinement by decomposing the Yang-Mills field, and we investigate the Polyakov loop average, static quark potential, chromoelectric flux,
 and induced monopole current for both Yang-Mills field and decomposed restricted field in both confinement and deconfinement phase at finite temperature.
We further discuss the role of the chromomagnetic monopole in the confinement/deconfinement phase transition.
}
\FullConference{XIII Quark Confinement and the Hadron Spectrum - Confinement2018\\
		31 July - 6 August 2018\\
		Maynooth University, Ireland}

\ifx\pdfoutput\relax\let\pdfoutput=\undefined\fi
\newcount\msipdfoutput
\ifx\pdfoutput\undefined\else
\ifcase\pdfoutput\else
\ifx\paperwidth\undefined\else
\ifdim\paperheight=0pt\relax\else\pdfpageheight\paperheight\fi
\ifdim\paperwidth=0pt\relax\else\pdfpagewidth\paperwidth\fi
\fi\fi\fi
\ifx\msidvipdf\msiundefined\else
\AtBeginDvi{}
\fi
\begin{document}
\section{Introduction}

The dual superconductivity is a promising mechanism for quark confinement
\cite{dualsuper}. To establish the dual superconductivity picture, we must
show that magnetic monopoles play a dominant role in quark confinement. For
this purpose, we have constructed a new framework for the $SU(N)$ Yang-Mills
theory on the lattice, \emph{called the decomposition method}, \emph{ }which
gives the decomposition of a gauge link variable $U_{x,\mu}=X_{x,\mu}V_{x,\mu
}$ to extract a variable $V_{x,\mu}$ called the restricted field as the
dominant mode for quark confinement in the gauge independent way. (See
\cite{KKSS15} for a review.) This formulation can overcome criticism raised
for the Abelian projection method \cite{tHooft81} for extracting
\textit{Abelian magnetic monopoles}, i.e., the magnetic monopole is obtained
only in special Abelian gauges such as the maximal Abelian (MA) gauge
\cite{KLSW87}. The Abelian projection is nothing but a gauge fixing to break
the gauge symmetry, which breaks also the color symmetry (global symmetry).

In the new framework, the $SU(3)$ Yang-Mills theory has two options for
choosing the fundamental field variables: the minimal and maximal options.
These two options are discriminated by the maximal stability subgroup
$\tilde{H}$ of the gauge group $SU(3)$. In the minimal option, the maximal
stability group is a non-Abelian group $\tilde{H}=U(2)$ and the restricted
field is used to extract non-Abelian magnetic monopoles. The minimal option is
suggested from the non-Abelian Stokes theorem for the Wilson loop operator in
the fundamental representation. In the preceding works, we have provided
numerical evidences of the non-Abelian dual superconductivity using the
minimal option for the $SU(3)$ Yang-Mills theory on a lattice. In the maximal
option, the maximal stability group is an Abelian group $\tilde{H}=U(1)\times
U(1)$, the maximal torus subgroup of $SU(3)$. This decomposition was first
constructed by Cho and Faddeev and Niemi \cite{SUN-decomp} by extending the
Cho-Duan-Ge-Faddeev-Niemi (CDGFN) decomposition for the $SU(2)$ case
\cite{CFNS-C}, and is nothing but the gauge invariant extension of the Abelian
projection in the maximal Abelian gauge. Therefore, the restricted field in
the maximal option involves only the Abelian magnetic monopole.

In this talk, we investigate the confinement/deconfinement phase transition at
finite temperature from a viewpoint of the dual superconductivity in the both
minimal and maximal options. The preliminary results are given by preceding
works \cite{lattice2013-16}\cite{SCGT15}. For this purpose, we examine several
quantities constructed by the restricted fields in both options as well as the
original Yang-Mills field at finite temperature, e.g., the distribution and
average of the Polyakov loops, the static potential from the Wilson loop, the
dual Meissner effects, and so on.\ The the dual Meissner effects at finite
temperature\ can be examined by measuring the distribution of the
chromo-electric field strength (or chromo flux) generated from a pair of the
static quark and antiquark and the associated magnetic-monopole current
induced around it. In particular, we discuss the role of the non-Abelian
magnetic monopole in confinement/deconfinement phase transition.

\section{Gauge link decompositions}

In this section, we give a brief review of \emph{the decomposition method},
which enables one to extract the dominant mode for quark confinement in the
$SU(N)$ Yang-Mills theory (see \cite{KKSS15} in detail). We decompose the
gauge link variable $U_{x,\mu}$ into the product of the two variables,
$V_{x,\mu}$ and $X_{x,\mu}$, in such a way that the new variable $V_{x,\mu}$
is transformed by the full $SU(N)$ gauge transformation $\Omega_{x}$ as the
gauge link variable $U_{x,\mu}$, while $X_{x,\mu}$ transforms as the site
variable:
\begin{subequations}
\label{eq:gaugeTransf}%
\begin{align}
U_{x,\mu}  &  =X_{x,\mu}V_{x,\mu}\in G=SU(N),\\
U_{x,\mu}  &  \longrightarrow U_{x,\nu}^{\prime}=\Omega_{x}U_{x,\mu}%
\Omega_{x+\mu}^{\dag},V_{x,\mu}\longrightarrow V_{x,\nu}^{\prime}=\Omega
_{x}V_{x,\mu}\Omega_{x+\mu}^{\dag},\text{ \ }X_{x,\mu}\longrightarrow
X_{x,\nu}^{\prime}=\Omega_{x}X_{x,\mu}\Omega_{x}^{\dag}.
\end{align}
From the physical point of view, $V_{x,\mu}$ could be the dominant mode for
quark confinement, while $X_{x,\mu}$ is the remainder part. For the $SU(3)$
Yang-Mills theory, we have two possible options discriminated by the stability
subgroup of the gauge group, which we call the minimal and maximal options.

\subsection{Minimal option}

The minimal option is obtained for the stability subgroup $\tilde
{H}=U(2)=SU(2)\times U(1)$. By introducing a single color field,
$\bm{h}_{x}=\xi_{x}\frac{\lambda^{8}}{2}\xi_{x}^{\dag}\in Lie[SU(3)/U(2)]$,
with $\lambda^{8}$ being the last diagonal Gell-Mann matrix and $\xi_{x}$ an
$SU(3)$ group element, the decomposition is obtained by solving the defining
equations:
\end{subequations}
\begin{equation}
D_{\mu}^{\epsilon}[V]\bm{h}_{x}:=\frac{1}{\epsilon}\left[  V_{x,\mu
}\bm{h}_{x+\mu}-\bm{h}_{x}V_{x,\mu}\right]  =0\text{ }. \label{eq:def1-min}%
\end{equation}
The defining equation can be solved exactly, and the solution is given by
\begin{subequations}
\label{eq:decomp-min}%
\begin{align}
X_{x,\mu}  &  =\widehat{L}_{x,\mu}^{\dag}\det(\widehat{L}_{x,\mu})^{1/3}%
g_{x}^{-1},\text{ \ }V_{x,\mu}=X_{x,\mu}^{\dag}U_{x,\mu}=g_{x}\widehat
{L}_{x,\mu}U_{x,\mu},\\
\widehat{L}_{x,\mu}  &  :=\left(  L_{x,\mu}L_{x,\mu}^{\dag}\right)
^{-1/2}L_{x,\mu},\text{ \ \ }L_{x,\mu}:=\frac{5}{3}\mathbf{1}+\frac{2}%
{\sqrt{3}}(\bm{h}_{x}+U_{x,\mu}\bm{h}_{x+\mu}U_{x,\mu}^{\dag})+8\bm{h}_{x}%
U_{x,\mu}\bm{h}_{x+\mu}U_{x,\mu}^{\dag}.
\end{align}
Here, the variable $g_{x}:=e^{i2\pi q/3}\exp(-ia_{x}^{0}\bm{h}_{x}%
-i\sum\nolimits_{j=1}^{3}a_{x}^{(j)}\mathbf{u}_{x}^{(j)})$ is the $U(2)$ part
which is undetermined from eq(\ref{eq:def1-min}) alone, $\mathbf{u}_{x}^{(j)}$
's are $su(2)$-Lie algebra valued, and \thinspace$q$ is an integer. Note that
the above defining equation with $g_{x}=\mathbf{1}$ corresponds to the
continuum version: $D_{\mu}[\bm{V}]\bm{h}(x)=0$ and $\mathrm{tr}(\bm{X}_{\mu
}(x)\bm{h}(x))$ $=0$. In the continuum limit, indeed, the decomposition in the
continuum theory is reproduced.

The decomposition is uniquely obtained as the solution of the defining
equation, once a set of color fields $\ \{\bm{h}_{x}\}$ are given. To
determine the configuration of color fields, we use the reduction condition of
minimizing the functional:
\end{subequations}
\begin{equation}
F_{\text{red}}[\bm{n}_{x}^{(8)}]=\sum_{x,\mu}\mathrm{tr}\left\{  (D_{\mu
}^{\epsilon}[U_{x,\mu}]\bm{h}_{x})^{\dag}(D_{\mu}^{\epsilon}[U_{x,\mu
}]\bm{h}_{x})\right\}  . \label{eq:reduction-min}%
\end{equation}

\subsection{Maximal option}

The maximal option is obtained for the stability subgroup of the maximal torus
subgroup of $G$: $\tilde{H}=U(1)\times U(1)$. By introducing the color field,
$\bm{n}_{x}^{(3)}=\xi_{x}\frac{\lambda^{3}}{2}\xi_{x}^{\dag}\in
Lie[SU(3)/U(1)\times U(1)],\quad\ \bm{n}_{x}^{(8)}=\xi_{x}\frac{\lambda^{8}%
}{2}\xi_{x}^{\dag}\in Lie[SU(3)/U(2)],$ with $\lambda^{3}$, $\lambda^{8}$
being the two diagonal Gell-Mann matrices and $\xi$ an $SU(3)$ group
element,\ the decomposition is obtained by solving the defining equations:
\begin{equation}
D_{\mu}^{\epsilon}[V]\bm{n}_{x}^{(j)}:=\frac{1}{\epsilon}\left[  V_{x,\mu
}\bm{n}_{x+\mu}^{(j)}-\bm{n}_{x}^{(j)}V_{x,\mu}\right]  =0\quad(j=3,8)\text{
.} \label{eq:def-max-1}%
\end{equation}
The defining equation can be solved exactly, and the solution is given by
\begin{subequations}
\label{eq:decomp_max}%
\begin{align}
X_{x,\mu}  &  =\widehat{K}_{x,\mu}^{\dag}\det(\widehat{K}_{x,\mu})^{1/3}%
g_{x}^{-1},\text{\ \ }V_{x,\mu}=X_{x,\mu}^{\dag}U_{x,\mu},\\
\widehat{K}_{x,\mu}  &  :=\left(  K_{x,\mu}K_{x,\mu}^{\dag}\right)
^{-1/2}K_{x,\mu},\text{ \ \ }K_{x,\mu}:=\mathbf{1}+6(\bm{n}_{x}^{(3)}U_{x,\mu
}\bm{n}_{x+\mu}^{(3)}U_{x,\mu}^{\dag})+6(\bm{n}_{x}^{(8)}U_{x,\mu
}\bm{n}_{x+\mu}^{(8)}U_{x,\mu}^{\dag}).
\end{align}
Here, the variable $g_{x}:=e^{i2\pi q/3}\exp(-ia_{x}^{3}\bm{n}_{x}%
^{(3)}-ia_{x}^{(8)}\bm{n}_{x}^{(8)})$ with integer $q$ is the $U(1)\times
U(1)$ part which is undetermined from eq(\ref{eq:def-max-1}) alone. Note that
the above defining equation with $g_{x}=\mathbf{1}$ corresponds to the
continuum version: $D_{\mu}[\bm{V}]\bm{n}^{(j)}(x)=0$ and $\mathrm{tr}%
(\bm{X}_{\mu}(x)\bm{n}^{(j)}(x))$ $=0$. In the continuum limit, we can
reproduce the decomposition in the continuum theory.

The decomposition is uniquely obtained as the solution (\ref{eq:reduction-max}%
) of the defining equations, once a set of color fields $\left\{
\bm{n}_{x}^{(3)},\bm{n}_{x}^{(8)}\right\}  $ are given. To determine the
configuration $\left\{  \bm{n}_{x}^{(3)},\bm{n}_{x}^{(8)}\right\}  $ of color
fields, we use the reduction condition of minimizing the functional:%
\end{subequations}
\begin{equation}
F_{\text{red}}[\bm{n}_{x}^{(3)},\bm{n}_{x}^{(8)}]=\sum_{x,\mu}\sum
_{j=3,8}\mathrm{tr}\left\{  (D_{\mu}^{\epsilon}[U_{x,\mu}]\bm{n}_{x}%
^{(j)})^{\dag}(D_{\mu}^{\epsilon}[U_{x,\mu}]\bm{n}_{x}^{(j)})\right\}  .
\label{eq:reduction-max}%
\end{equation}
Note that, the resulting decomposition is the gauge-invariant extension of the
Abelian projection in the maximal Abelian (MA) gauge.

\section{Numerical simulations on the lattice}

We set up the numerical simulations at finite temperature adopting the
standard Wilson action with the inverse gauge coupling constant $\beta
=2N_{c}/g^{2}$ ($N_{c}=3)$ and using the pseudo heat-bath algorithm and the
over-relaxation algorithm to generate the gauge field configurations (link
variables) $\{U_{x,\mu}\}$ on the lattice of size $N_{s}^{3}\times N_{T}$ . We
prepare 1000 gauge configurations every 100 seeps with 8 over relaxations
after 8000 thermalization sweeps with cold start for the fixed spatial size
$N_{s}$ and the temporal size $N_{T}$: $N_{s}=24$, $N_{T}=6$, where the
temperature varies by changing the coupling $\beta$ ($5.75\leq\beta\leq6.50$).

We obtain the color field configurations for the minimal and maximal options
by solving the reduction conditions, and then we perform the decomposition of
the gauge link variable $U_{x,\mu}=X_{x,\mu}V_{x,\mu}$ by using the formula
given in the previous section, i.e., for the minimal option
eq(\ref{eq:decomp-min}) \ with the color field$\ \{\bm{h}_{x}\}$ by minimizing
eq(\ref{eq:reduction-min}), for the maximal option eq(\ref{eq:decomp_max}) and
the color field $\left\{  \bm{n}_{x}^{(3)},\bm{n}_{x}^{(8)}\right\}  $ by
minimizing eq(\ref{eq:reduction-max}). In the measurement of the Polyakov loop
average and the Wilson loop average defined below, we apply the APE smearing
technique \cite{Albanese87} to reduce noises .

\subsection{Polyakov-loop average at the confinement/deconfinement transition}

First, we investigate the distribution of single Polyakov loops. For a set of
the original gauge field configurations $\{U_{x,\mu}\}$ and the restricted
gauge field configurations $\{V_{x,\mu}\}$ in the minimal and maximal options,
we define the respective Polyakov loops by
\begin{equation}
P_{U}^{\mathrm{YM}}(\mathbf{x}):=\frac{1}{3}\text{\textrm{tr}}\left(
P\prod\limits_{t=1}^{N_{T}}U_{(\mathbf{x},t),4}\right)  ,\ P_{V}^{\min
}(\mathbf{x}):=\frac{1}{3}\text{\textrm{tr}}\left(  P\prod\limits_{t=1}%
^{N_{T}}V_{(\mathbf{x},t),4}^{(\min)}\right)  ,\ P_{V}^{\max}(\mathbf{x}%
):=\frac{1}{3}\text{\textrm{tr}}\left(  P\prod\limits_{t=1}^{N_{T}%
}V_{(\mathbf{x},t),4}^{(\max)}\right)  \text{ ,} \label{eq;Polyakov-loop}%
\end{equation}
and the space-averaged Polyakov loops by
\begin{equation}
P_{YM}:=\frac{1}{L^{3}}\sum_{\mathbf{x}}P_{U}^{\mathrm{YM}}(\mathbf{x}%
),\ \ \ P_{\min}:=\frac{1}{L^{3}}\sum_{\mathbf{x}}P_{V}^{\min}(\mathbf{x}%
),\ P_{\max}:=\frac{1}{L^{3}}\sum_{\mathbf{x}}P_{V}^{\max}(\mathbf{x}),
\label{eq:Ployakov-loop-ave-S}%
\end{equation}
where the value of the Polyakov loop is averaged over the space volume for
each configuration.%

\begin{figure}[hbt] \centering
\includegraphics
[height=42mm, clip, viewport=0  5 360 250 ]{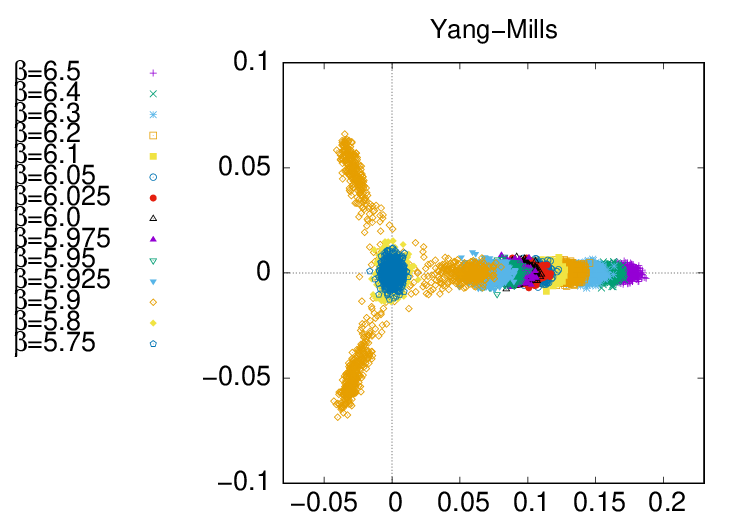}
\
\includegraphics
[height=42mm, clip, viewport=100 5 360 250,]{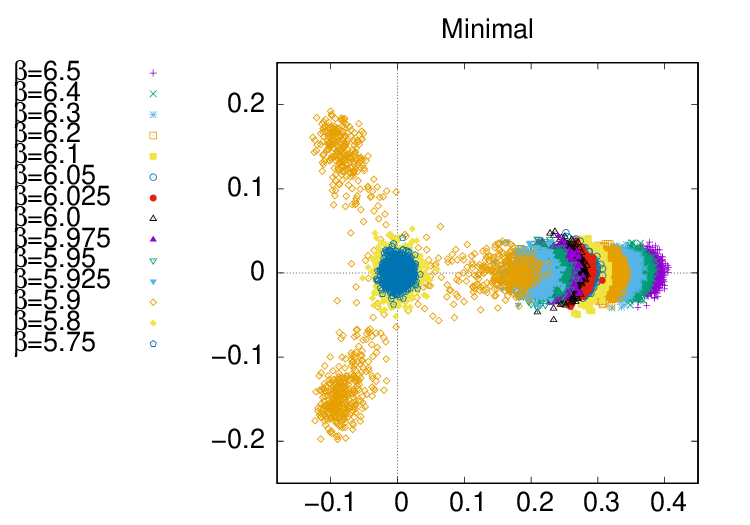}
\includegraphics
[height=42mm, clip, viewport=100 5 360 250 ]{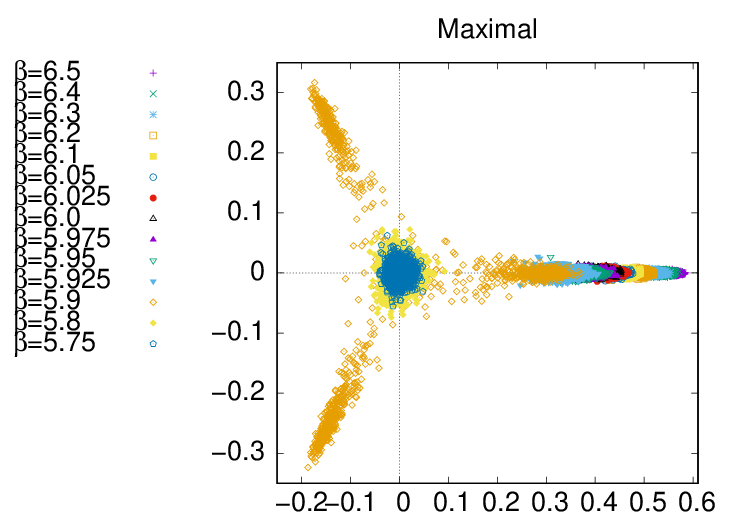}
\caption{
The distribution of the space-averaged Polyakov loop  on the complex plane:
(Left) Yang-Mills field (Middlel) restricted field in minimal option
(Right) restricted field in maximal option.
}\label{fig:PLPdist}%
\end{figure}%

Figure \ref{fig:PLPdist}\thinspace shows the plots of and $\{P_{U}%
^{\mathrm{YM}}(\mathbf{x})\}$, $\{P_{V}^{\min}(\mathbf{x})\}$ and
$\{P_{V}^{\max}(\mathbf{x})\}$ on the complex plane measured from a set of the
original Yang-Mills field configurations and a set of the restricted field
configurations in the minimal and maximal options, respectively. We find that
the value of the space-averaged Polyakov loops are different option by option,
but all the distributions on the complex plane equally reflect the expected
center symmetry $Z(3)$ of the $SU(3)$ gauge group.%

\begin{figure}[tbp] \centering
\includegraphics[height=48mm]{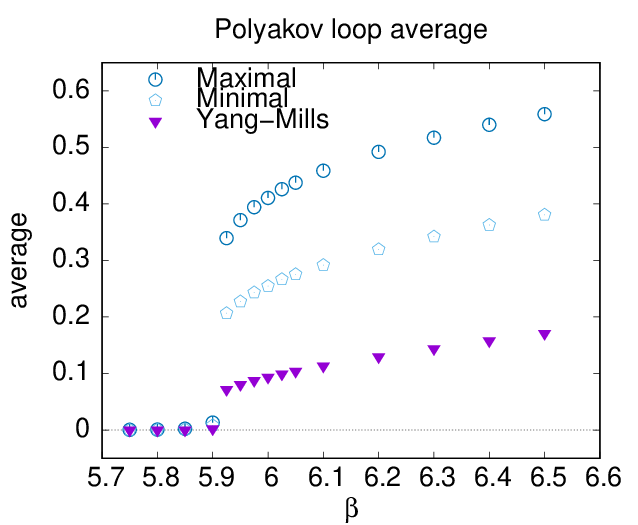}
\ \ \ \ \
\includegraphics[height=48mm]{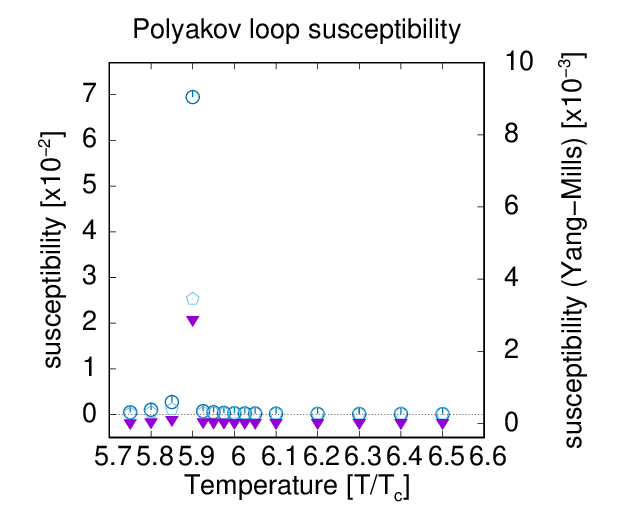}
\caption{
(Left) Polyakov-loop averages for the original field and the restricted field in the minimal
and maximal options from bottom to top.
(Right) The susceptibility of the Polyakov loop for the original field
and the restricted field in the minimal and maximal options from bottom to top.
}\label{fig:PLP_ave}%
\end{figure}%

The left panel of Figure \ref{fig:PLP_ave} shows the comparison of \ three
Polyakov-loop averages, i.e., $\left\langle P_{YM}\right\rangle $,
$\left\langle P_{\min}\right\rangle $, and $\left\langle P_{\max}\right\rangle
$, for various temperature ($\beta$), where the symbol $\left\langle
\mathcal{O}\right\rangle $ denotes the average of the operator $\mathcal{O}$
over the ensemble of the configurations. The right panel of Fig.
\ref{fig:PLP_ave} shows their susceptibilities, $\chi:=\left\langle P_{\ast
}^{2}\right\rangle -\left\langle P_{\ast}\right\rangle ^{2}$, with $P_{\ast}$
being one of $P_{YM}$, $P_{\min}$, and $P_{\max}$. These panels clearly show
that both the minimal and maximal options reproduce the critical point of the
original Yang-Mills field theory. Thus, these three Polyakov-loop averages
give the identical critical temperature, i.e., $\beta=5.9$.

\subsection{Static quark--antiquark potential at finite temperature}

\begin{figure}[bt] \centering
\includegraphics[height=28mm]
{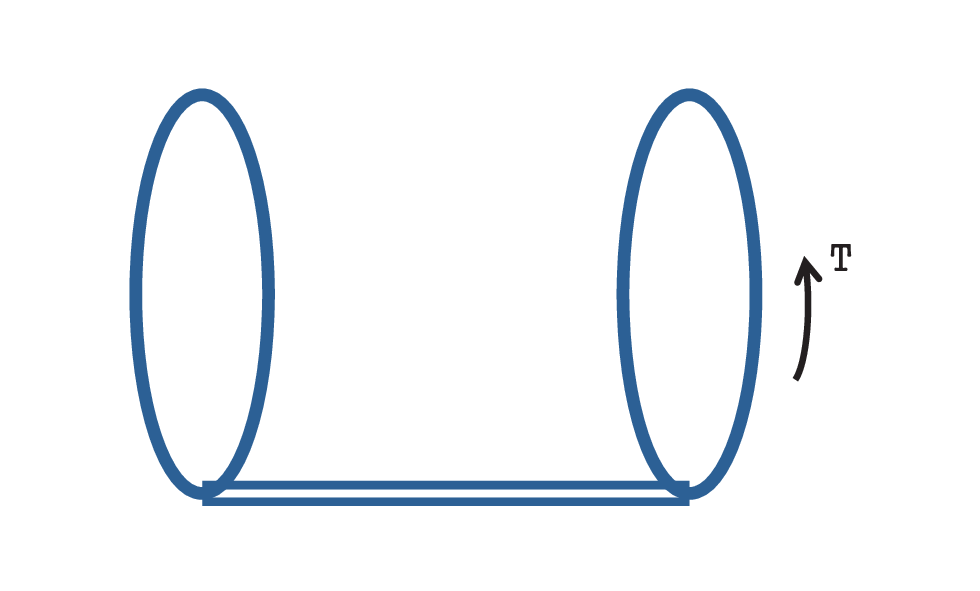}
\ \ \ \ \
\includegraphics[height=31mm]{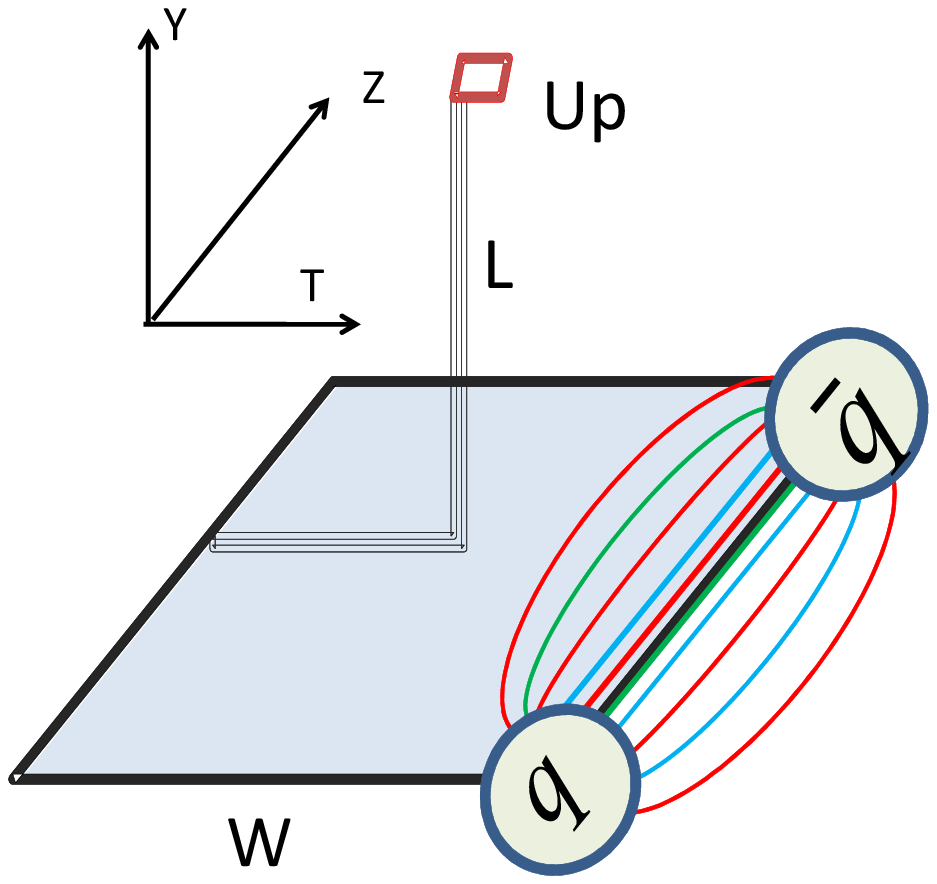}
\ \ \ \ \
\includegraphics[height=28mm]{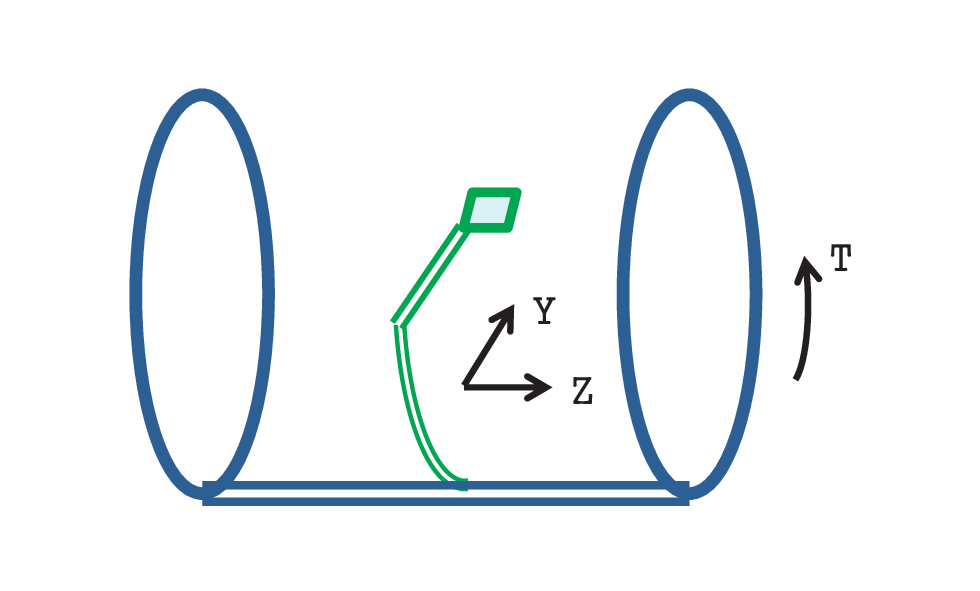}
\caption{
The set up for the measurement of the static quark-antiquark potential at finite temperature.
(Left) the Wilson loop
(Middle) The gauge-invariant operator $tr(WLU_{p}L^{õ})$  between a plaquette $U_{p}$  and the Wilson loop $W$.
(Right) Measurement of the chromo-flux at finite temperature.
}\label{fig:Wilson-loop}%
\end{figure}

Next, we investigate the static quark--antiquark potential at finite
temperature. To obtain the static potential at finite temperature, we adopt
the Wilson loop operator, (see the left panels of Fig.\ref{fig:Wilson-loop}),
which is defined for the rectangular loop $C$ with the spatial length $R$ and
the temporal length $\tau$ which is maximally extended in the temporal
direction, i.e., $\tau=1/T$. According to the standard argument, for large
$\tau$ i.e., small $T$, the static potential is obtained from the original
gauge field $U$ and the restricted gauge field $V$:%
\begin{equation}
V(R;U):=-\frac{1}{\tau}\log\left\langle W_{U}\right\rangle =-T\log\left\langle
W_{U}\right\rangle ,\quad V(R;V):=-\frac{1}{\tau}\log\left\langle
W_{V}\right\rangle =-T\log\left\langle W_{V}\right\rangle .
\label{eq:PL-potentials}%
\end{equation}
%

%

\begin{figure}[tbp] \centering
\includegraphics
[height=46mm, clip, viewport=0 0 360 250 ]{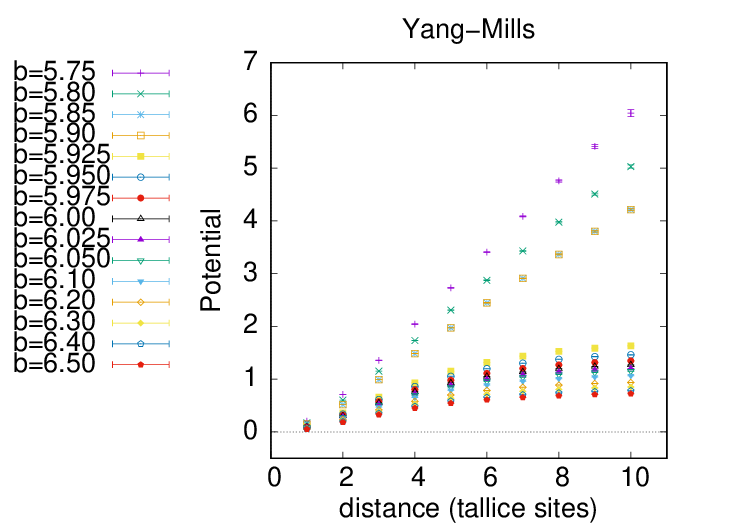}%
\includegraphics
[height=46mm,clip,viewport=95 0 360 250]{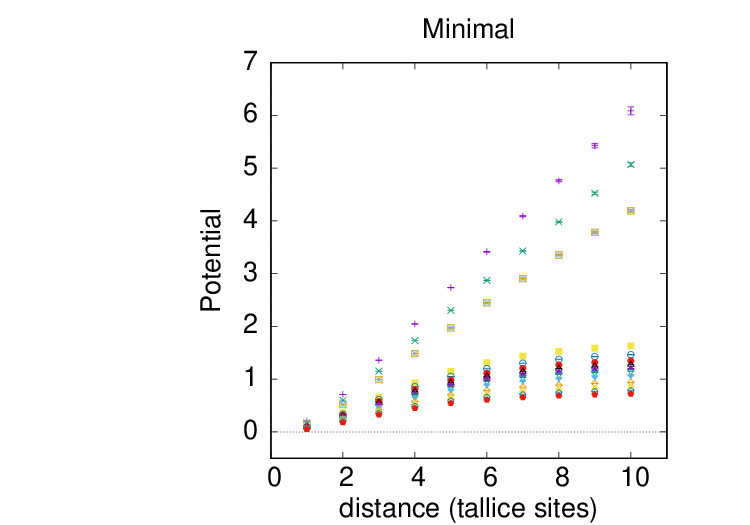}%
\includegraphics
[height=46mm, clip, viewport=95 0 360 250]{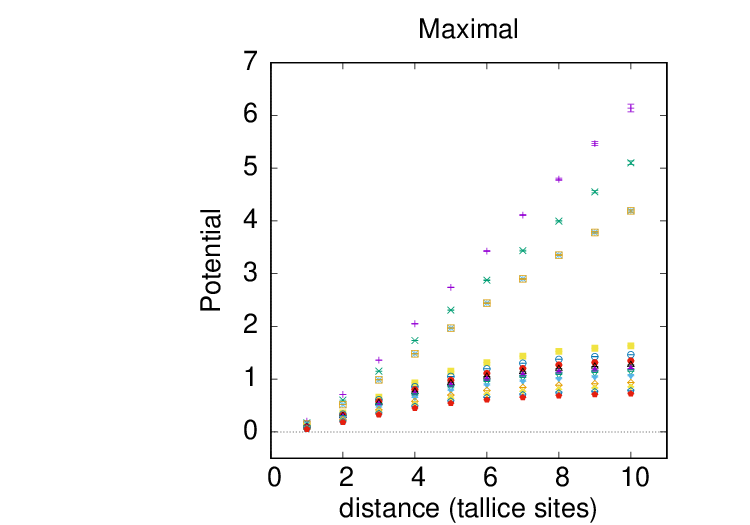}
\caption{The static quark potential obtained using the maximally extended Wilson loop from the data
set (II) at various temperatures $N_{T}=6$ and $ 5.75 \le \beta \le 6.5$
(Left) original gauge field,
(Middle) restricted field in the minimal option,
(Right) restricted field in the maximal option.
(applying APE smearing 8 times with weight 0.2)
} \label{fig:wloop-T-2}%
\end{figure}%

Figure \ref{fig:wloop-T-2} shows the static potentials at various temperatures
($\beta$) calculated according to the definition (\ref{eq:PL-potentials}). By
comparing these results, we find that the potential is reproduced by the
restricted field alone in both options. Therefore, we have shown the
restricted $V$-field dominance for both options in the static potential at
finite temperature.

\subsection{Chromo-flux tube at finite temperature}

We proceed to investigate the non-Abelian dual Meissner effect at finite
temperature. For this purpose, we measure the chromo-flux created by a
quark-antiquark pair, which is represented by the maximally extended Wilson
loop $W$ as given in the middle and right panel of Fig.\ref{fig:Wilson-loop}.
The chromo-field strength, i.e., the field strength of the chromo-flux created
by the Wilson loop $W$ as the source, is measured by using a plaquette
variable $U_{p}$ as the probe operator for the field strength. We use the
gauge-invariant correlation function which is the same as that used at zero
temperature \cite{GMO90}:
\begin{equation}
F_{\mu\nu}^{q\bar{q}}=\sqrt{\frac{\beta}{6}}\rho_{_{U_{P}}}\text{, \ \ \ }%
\rho_{_{U_{P}}}:=\frac{\left\langle \mathrm{tr}\left(  WLU_{p}L^{\dag}\right)
\right\rangle }{\left\langle \mathrm{tr}\left(  W\right)  \right\rangle
}-\frac{1}{N_{c}}\frac{\left\langle \mathrm{tr}\left(  U_{p}\right)
\mathrm{tr}\left(  W\right)  \right\rangle }{\left\langle \mathrm{tr}\left(
W\right)  \right\rangle }\text{ }, \label{eq:Op}%
\end{equation}
where $L$ is the Wilson line connecting the source $W$ and the probe $U_{p}$
to guarantee the gauge-invariance.
Indeed, in the naive continuum limit, the connected correlator $\rho_{_{U_{P}%
}}$ reduces to $\rho_{_{U_{P}}}\overset{\varepsilon\rightarrow0}{\simeq
}g\epsilon^{2}\left\langle \mathcal{F}_{\mu\nu}\right\rangle _{q\bar{q}%
}:=\frac{\left\langle \mathrm{tr}\left(  g\epsilon^{2}\mathcal{F}_{\mu\nu
}L^{\dag}WL\right)  \right\rangle }{\left\langle \mathrm{tr}\left(  W\right)
\right\rangle }+O(\epsilon^{4})$.%

\begin{figure}[tbp] \centering
\includegraphics[height=38mm ]
{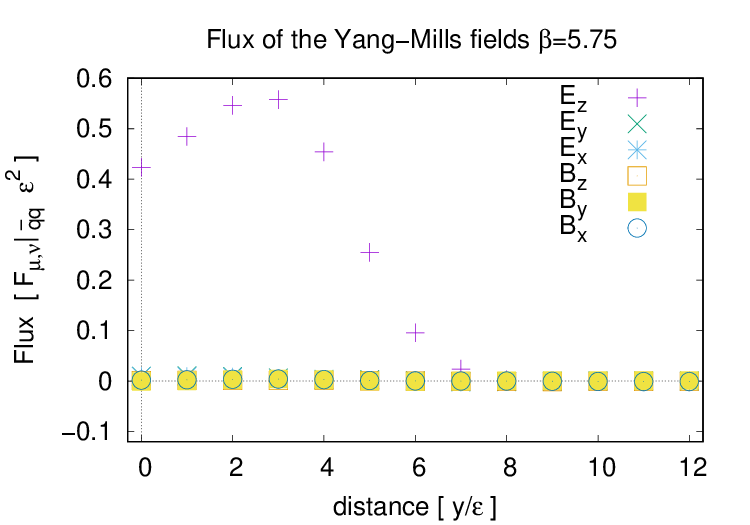}\includegraphics[height=38mm ]
{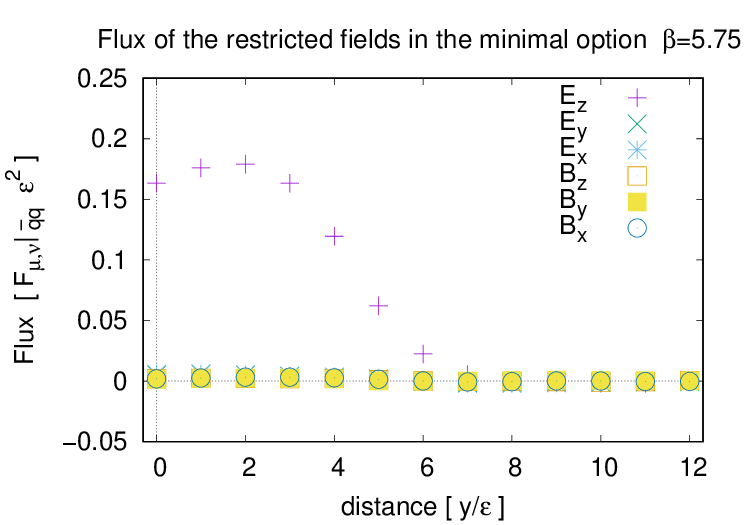}\includegraphics[height=38mm ]
{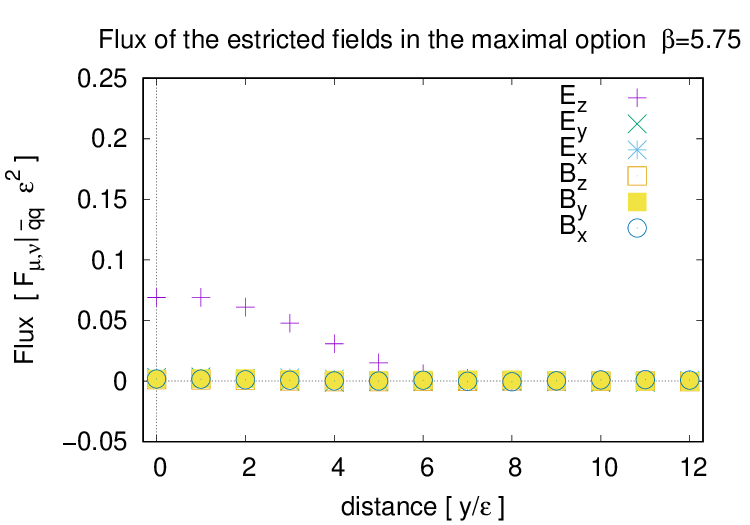}%
\\

\includegraphics[height=38mm ]
{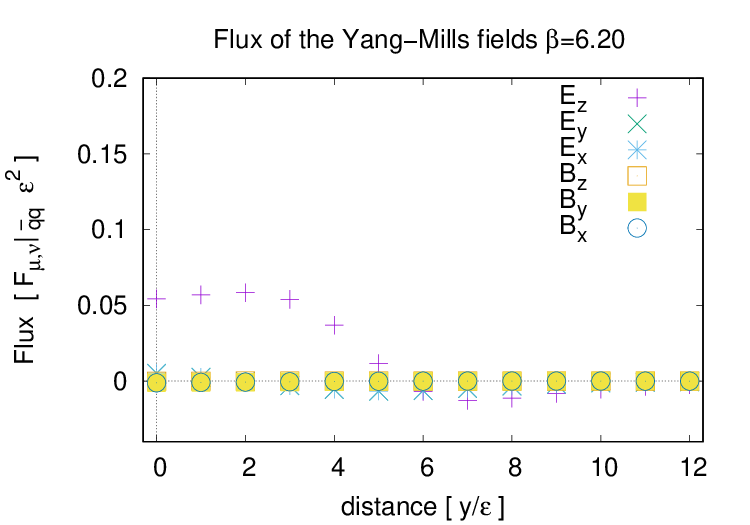}\includegraphics[height=38mm ]
{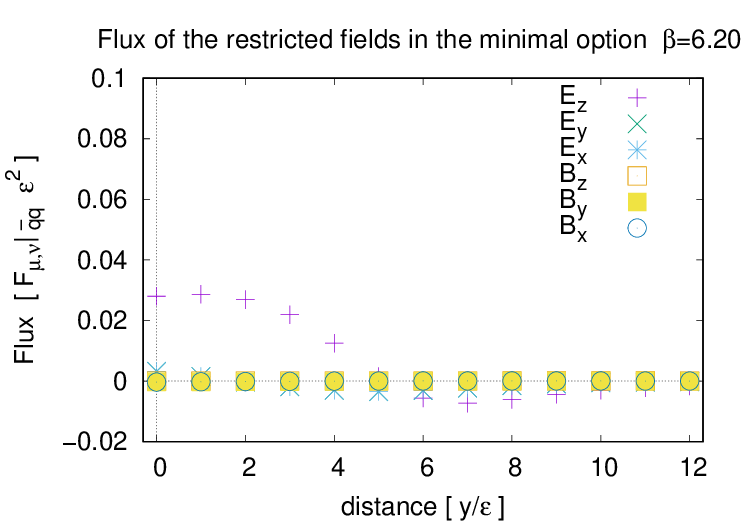}\includegraphics[height=38mm ]
{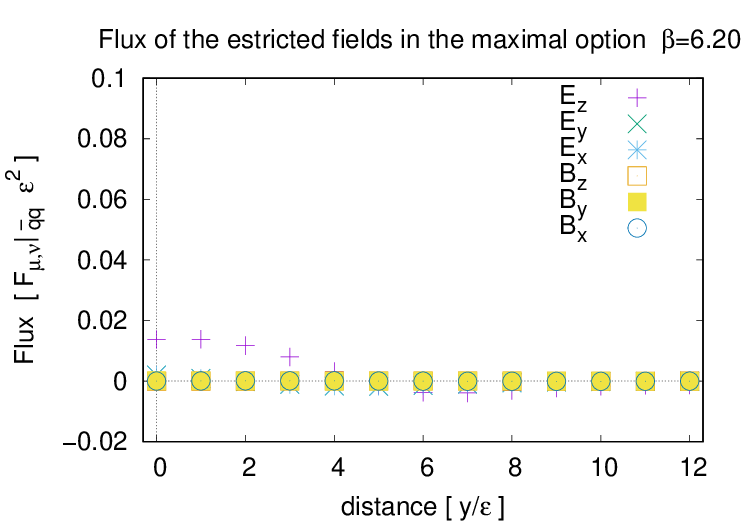}\caption{
Measurement of chromo flux: $E_x=F^{q\bar{q}}_{14}$, $E_y=F^{q\bar{q}}_{24}$, $E_Z=F^{q\bar{q}}_{34}$,
$B_x=F^{q\bar{q}}_{23}$, $B_y=F^{q\bar{q}}_{31}$, $B_z=F^{q\bar{q}}_{12}$  .
The upper panels show the plots of  chromo fluxes at a low temperature in the confiment phase  ($T < T_c$),
and the lower panels show those  at a high temperature in the deconfinement phase .($T_c <  T$).
The left-column panels show the chromo flux for Yang-Mills field, and  the middle- and right-column  panels
show the cheomo flux for the restricted field in the minimal and mxaimal options, respectively
.}\label{fig:flux-phase}%
\end{figure}%

Figure \ref{fig:flux-phase} shows the chromo flux measured by using
eq(\ref{eq:Op}). At a low-temperature in the confinement phase, $T<T_{c}$, we
observe that only the component $E_{z}$ of the chromoelectric flux tube in the
direction connecting a quark and antiquark pair is non-vanishing, while the
other components take vanishing values. (See the upper panels of Fig.
\ref{fig:flux-phase}. ) At a high-temperature in the deconfinement phase,
$T>T_{c}$, we observe the non-vanishing component $E_{y}$ orthogonal to the
chromoelectric flux, which means no more squeezing of the chromoelectric flux
tube. (See the lower panels of Fig. \ref{fig:flux-phase} .) This is a
numerical evidence for the disappearance of the dual Meissner effect in the
high-temperature deconfinement phase.

\subsection{Magnetic--monopole current and dual Meissner effect at finite
temperature}%

\begin{figure}[tbp] \centering
\includegraphics[height=47mm, clip, viewport= 0 0 360 252  ]{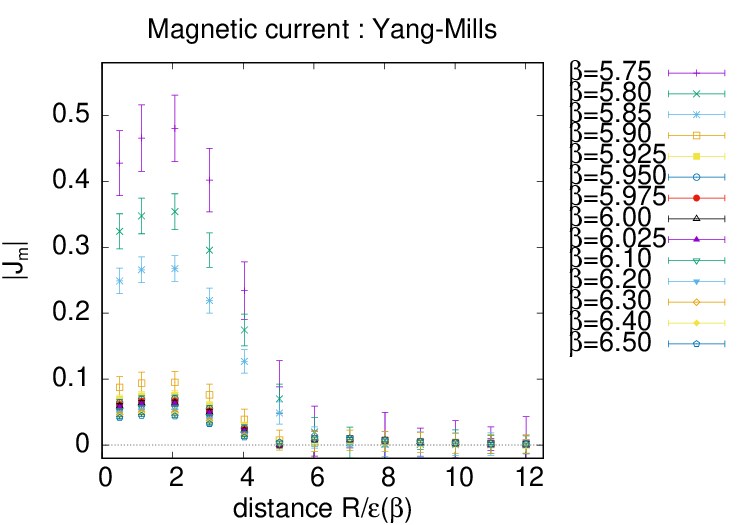}\includegraphics[height=47mm, clip, viewport= 20 0 270 252 ]{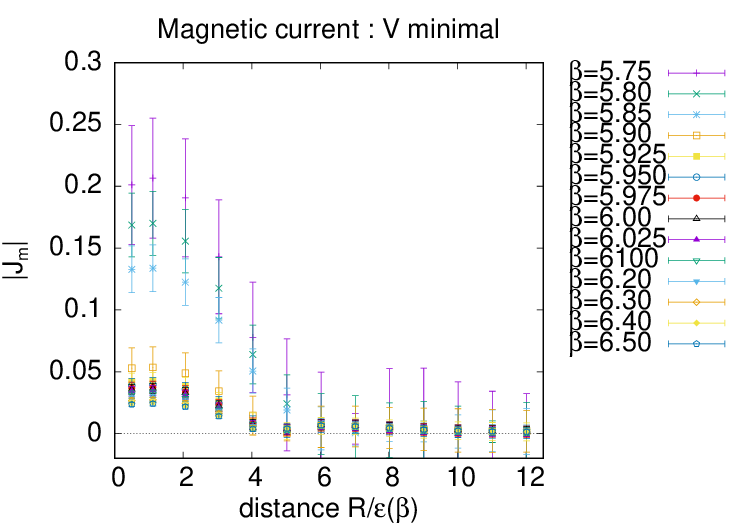}\includegraphics[height=47mm, clip, viewport= 20 0 270 252 ]{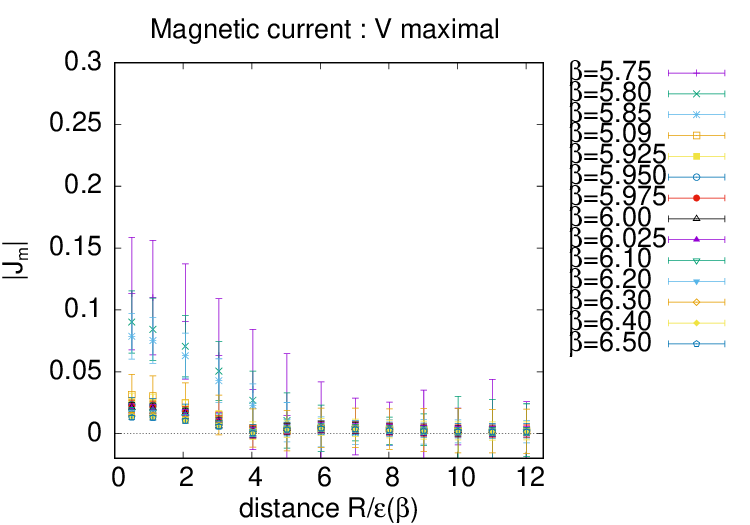}\caption{
The magnitude $\sqrt{k_\mu k_\mu} $ of the induced magnetic current $k_{\mu}$ around the flux tube connecting
the quark-antiquark pair as a function of the distance y from the z axis for various values of $\beta$ i.e.,
temperature. (Left) Yang-Mills filed (Middle) minimal option, (Right) maximal option.
}\label{fig:m_current}%
\end{figure}%

Finally, we investigate the dual Meissner effect by measuring the
magnetic--monopole current $k$ induced around the chromo-flux tube created by
the quark-antiquark pair. We use the magnetic--monopole current $k$ defined
by
\begin{equation}
k_{\mu}(x)=\frac{1}{2}\epsilon_{\mu\nu\alpha\beta}\left(  F_{\alpha\beta
}^{q\bar{q}}[V](x+\hat{\nu})-F_{\alpha\beta}^{q\bar{q}}[V](x)\right)  ,
\label{eq:m_current}%
\end{equation}
where $F[V]$ is the field strength of the restricted field $V$. \ This
definition satisfies the conserved current, i.e., $\partial_{\mu}k_{\mu
}(x):=\sum_{\mu}\left(  k_{\mu}(x+\hat{\mu})-k_{\mu}(x)\right)  \equiv0$. Note
that the magnetic--monopole current (\ref{eq:m_current}) must vanish due to
the Bianchi identity as far as there exists no singularity in the gauge
potential,\ since the field strength\ is written by using differential
forms\ as $F[V]=dV,$ and then the magnetic--monopole current vanishes, i.e.,
$k:={}^{\ast}dF={}^{\ast}ddV=0$. We show that the magnetic--monopole current
defined in this way can be the order parameter for the
confinement/deconfinement phase transition, as suggested from the dual
superconductivity hypothesis. Fig.~\ref{fig:m_current} shows the result of the
measurements of the magnitude $\sqrt{k_{\mu}k_{\mu}}$ of the induced magnetic
current $k_{\mu}$ obtained according to (\ref{eq:m_current}) for various
temperatures ($\beta$). The current decreases as the temperature becomes
higher and eventually vanishes above the critical temperature for both
options. We observe the appearance and disappearance of the magnetic monopole
current in the low temperature phase and high temperature phase, respectively.


\section{Summary and outlook}

By using a new formulation of Yang-Mills theory, we have investigated two
possible options of the dual superconductivity at finite temperature in the
$SU(3)$ Yang-Mills theory, i.e., the Non-Abelian dual superconductivity in the
minimal and the maximal options which are to be compared with the conventional
Abelian dual superconductivity. In the measurement for both maximal and
minimal options as well as for the original Yang-Mills field at finite
temperature, we found the restricted $V$-field dominance in the string
tensions for both options.\ Then, we have investigated the dual Meissner
effect and found that the chromoelectric flux tube appears in both options in
the confining phase, but it disappears in the deconfinement phase. Thus both
options can be adopted as the low-energy effective description of the original
Yang-Mills theory. The detailed analysis will be appear in the subsequent
paper \cite{KKS2018}.

\subsection*{Acknowledgement}

This work is supported by Grant-in-Aid for Scientific Research (C) 24540252
and (C) 15K05042 from Japan Society for the Promotion Science (JSPS), and also
in part by JSPS Grant-in-Aid for Scientific Research (S) 22224003. The
numerical calculations are supported by the Large Scale Simulation Program of
High Energy Accelerator Research Organization (KEK): No.12-13(FY2012),
No.12/13-20(FY2012-13), No.13/14-23(FY2013-14), No.14/15-24(FY2014-15) and
No.16/17-20(FY2016-17) .

\end{document}